# Superconducting interaction charge in thallium-based high-$T_C$ cuprates: Roles of cation oxidation state and electronegativity

Dale R. Harshman [a,b*], Anthony T. Fiory [c]

[a] *Physikon Research Corporation, Lynden, WA 98264 USA, and*
[b] *Department of Physics, College of William and Mary, Williamsburg, VA 23187 USA*
[c] *Department of Physics, New Jersey Institute of Technology, Newark, NJ 07102 USA*

ABSTRACT

Superconductivity in the Tl-based cuprates encompasses a notably broad range of measured optimal transition temperatures $T_{C0}$, ranging from lowest in the charge-depleted Tl-1201 compounds ($Tl_{1-x}(Ba/Sr)_{1+y}La_{1-y}CuO_{5-\delta}$), such as $Tl_{0.7}LaSrCuO_5$ (37 K) and $TlBa_{1.2}La_{0.8}CuO_5$ (45.4 K), to highest in the Tl-1223 compound $TlBa_2Ca_2Cu_3O_{9\pm\delta}$ (133.5 K). Seven Tl-based cuprates are considered and compared using the model of superconductive pairing via electronic interactions between two physically separated charge reservoirs, where $T_{C0} \propto (\sigma\eta/A)^{1/2}\zeta^{-1}$ is determined by the superconducting interaction charge fraction $\sigma$, the number $\eta$ of $CuO_2$ layers, and the basal-plane area $A$, each per formula unit, and the transverse distance $\zeta$ between interacting layers. Herein it is demonstrated that $\sigma$ follows from the elemental electronegativity and the oxidation state of Tl, and other structurally analogous cations. The comparatively lower elemental electronegativity of Tl, in conjunction with its oxidation state, explains the higher $\sigma$ and $T_{C0}$ values in the Tl-based compounds relative to their Bi-based cuprate homologues. A derivation of $\sigma$ is introduced for the optimal $Tl_2Ba_2Ca_{\eta-1}Cu_\eta O_{2\eta+4}$ (for $\eta$ = 1, 2, 3) compounds, which exhibit a Tl oxidation state at or near +3, obtaining the fundamental value $\sigma_0$ = 0.228 previously established for $YBa_2Cu_3O_{6.92}$. Also reported is the marked enhancement in $\sigma$ associated with $Tl^{+1}$ and analogous inner-layer cations relative to higher-valence cations. For a model proposition of $\sigma = \sigma_0$, the fractional $Tl^{+1}$ content of the mixed-valence compound, $TlBa_2Ca_2Cu_3O_{9\pm\delta}$, is predicted to be 1/3 at optimization, in agreement with existing data. Charge depletion is illustrated for the two Tl-1201 compounds, where $\sigma < \sigma_0$ values are determined according to substitution of $Ba^{+2}$ or $Sr^{+2}$ by $La^{+3}$, and/or Tl depletion. Additionally, statistical analysis of calculated and experimental transition temperatures of 48 optimal superconductors shows an absence of bias in determining $\sigma$, $A$, and $\zeta$.

---

* Corresponding author at: Department of Physics, College of William and Mary, Williamsburg, VA 23187 USA.
*Email address:* drh@physikon.net (D. R. Harshman)

## 1. Introduction

The thallium-based high-$T_C$ cuprates, by virtue of having the widest variation in optimal transition temperature $T_{C0}$, encompassing a factor of 3.5, provide a rather unique opportunity for studying the systematic behavior of $T_{C0}$. Determined experimentally from the point of optimal doping, $T_{C0}$ can differ significantly between compounds with directly comparable layer components. Among optimal superconductors sharing the same number $\eta$ of $CuO_2$ layers per formula unit, $T_{C0}$ is found to vary over factors of 2.8, 2.5, and 1.3 for $\eta$ = 1, 2, and 3, respectively. A notable comparison, representative of the $\eta$ = 1 cuprates, can be made between $Bi_2(Sr_{1.6}La_{0.4})CuO_{6+\delta}$ and $Tl_2Ba_2CuO_6$ (denoted as 2201 compounds). Although belonging to different space groups, these materials share one-to-one correspondences between like structural layers, yet exhibit vastly different $T_{C0}$ values of 34 K [1-3] and 80 K [4], respectively; for $\eta$ = 2, $Bi_2Sr_2CaCu_2O_{8+\delta}$ and $Tl_2Ba_2CaCu_2O_8$ (2212 compounds) display measured $T_{C0}$ = 89 K [5] and 110 K [6], respectively, and this tendency continues for $\eta$ = 3 (2223 compounds). Data for $T_{C0}$ are cited after Ref. [7].

Such disparities in $T_{C0}$ between structural homologues illustrate the essential roles of crystal structure and the non-$CuO_2$ constituents in forming high-$T_C$ superconductive states. As detailed in prior work, it is essential to model high-$T_C$ compounds as layered structures comprising two types of charge reservoirs [7-9]. The non-$CuO_2$ structure, which commonly provides charge doping from metal cations associated with oxygen anions, is designated as type I and the structure containing $CuO_2$ layers is designated as type II. In this model, Coulomb interactions completely determine $T_{C0} \propto (\ell\zeta)^{-1}$, as expressed through the interaction-charge length $\ell \equiv (\sigma\eta/A)^{-1/2}$ and the interaction distance $\zeta$ given by the transverse distance between the two reservoirs' outer layers; $\sigma$ is the superconducting interaction charge fraction determined per type I interacting (outer) layer and $A$ is the basal-plane area, each given per formula unit [7].

This work focuses on calculating $\sigma$ by valency scaling methods, where a given composition is treated as cation substitutions in the type I reservoir relative to the exemplary compound $YBa_2Cu_3O_{6.92}$ having superconducting charge fraction $\sigma_0$ = 0.228. Valency scaling is shown to follow rules derived from the relative oxidation states and electronegativities of cations comprising the inner layers of the type I reservoir. Electronegativity of atoms, relative to that of Cu, is shown here to be important as it governs the efficacy of charge transfer along the hard axis; the level of interlayer charge transfer also affects charge allocation which is directly manifested in the value of $\sigma$. Apart from structural differences (mainly $\zeta$ and $A$), the aforementioned differences in $T_{C0}$ are thereby traced to several basic chemical attributes of the type I cations, in that the value $\sigma = \sigma_0$ is determined for $Tl_2Ba_2CuO_6$ and $Tl_mBa_2Ca_{\eta-1}Cu_\eta O_{2\eta+2+m}$ (m = 1, 2; $\eta$ = 2, 3), and the reduced values $\sigma \leq 0.5\ \sigma_0$ are determined for the Bi-based cuprates. Since $\sigma_0$ applies to a variety of disparate cuprates [7], it is considered to acquire a fundamental value.

A new application of the valency scaling method in this work is inclusion of the charge-depleted Tl-1201 compounds that were omitted from earlier treatments. Comprising a single $CuO_2$ layer and only one inner TlO layer per formula unit, compounds of the $Tl_{1-x}(Ba/Sr)_{1+y}La_{1-y}CuO_{5-\delta}$ system present the lowest transition temperatures of the Tl-based cuprate superconductors, correspond to $\sigma < \sigma_0$, and optimize via partial cation substitutions (e.g., Tl, Ba, Sr, and La) within the type I reservoir structure. The $TlBa_{1+y}La_{1-y}CuO_5$ compound, for example, is optimized at y ≈ 0.2 by partial substitution of $La^{+3}$ for $Ba^{+2}$, yielding $T_{C0}$ = 45.4 K, determined by averaging the magnetic susceptibility onset (45.5 K) and the resistivity midpoint (45.2 K) [10]. The related compound $Tl_{1-x}LaSrCuO_5$ is optimized by depleting the Tl content of the inner TlO layer with x = 0.36 ± 0.06 from site occupancy measurements; An x ≈ 0.3 sample exhibits resistive onset at 40 K, zero resistance at 37 K, and inductive onset at 37.5 K [11].

Further, the charge fraction $\sigma = \sigma_0$ for optimal single-TlO-layer compounds with $\eta$ = 2, 3 is shown to predict a mixture of +1 and +3 oxidation states of Tl with the fraction of +1 given by $f_{+1}$ = 1/3. Comparing the optimal $TlBa_2Ca_2Cu_3O_{9-\delta}$ with related non-optimal compositions produced by several cation substitutions and oxygen dopings, it is shown that experimental data for $f_{+1}$ increase monotonically to the theoretical 1/3 as the measured values of non-optimal $T_C$ increase to the optimal $T_{C0}$.

Numerous rationales exist for basing high-$T_C$ theory on electronic (i.e., non-lattice) pairing mechanisms [12], e.g., included in deductions from the quantum oscillations observed in overdoped $Tl_2Ba_2CuO_{6+\delta}$ [13]. Theoretical treatments of high-$T_C$ superconductivity focusing on temperature *vs.* doping phase diagrams are noted for being preferentially centered on the $CuO_2$ layers, considered either individually or in ensembles, and for viewing type I structures solely as vehicles for supplying doping charges to cuprate planes [12, 14, 15]. An approach of this kind is able to calculate $T_{C0}$ to the right order of magnitude [16]. Other electronic interaction models have proposed involvement of charges located outside the $CuO_2$ layers in application to $La_{2-x}(Ba,Sr)_xCuO_4$ and $YBa_2Cu_3O_7$ [17] and dynamic interlayer Coulomb interactions in analyzing $La_{1.85}Sr_{0.15}CuO_4$ [18].

In contrast with earlier investigations of possible correlations between $T_C$ and electronegativities [19-22], the approach presented herein places the focus on individual atoms. Specific to the cations in the type I inner layers, the additional influence of an element's electronegativity $\chi$ on fractional charge $\sigma$ is considered. Apart from the substitutional inner-layer cation, the surrounding local electronic (ionic) environment is essentially the same for applicable cuprates. Consequently, comparisons of the single (isolated) atom electronegativities are considered satisfactory for providing first-order estimations. Differences in electronegativities are shown to explain not only the equivalence of $\sigma$ among numerous cuprate compounds to $\sigma_0$ of $YBa_2Cu_3O_{6.92}$, but also why seemingly similar compounds can exhibit strikingly disparate optimal transition temperatures. Take, for example, the aforementioned $Bi_2Sr_2CaCu_2O_8$ and $Tl_2Ba_2CaCu_2O_8$. Both are 2212-cuprate compounds with type I structures containing cations of valences +3 in the inner layers and +2 in the outer layers. Apart from ionic size, the only significant electronic distinction is in the electronegativity; here, Bi possesses an elemental Pauling electronegativity of $\chi$ = 2.02, significantly above $\chi$ = 1.90 for Cu, while for Tl, $\chi$ = 1.62 and is substantially below that of Cu. Being electronegative relative to Cu, Bi tends to hold onto and attract electrons, preventing their transfer to the outer SrO layers, while the comparatively electropositive Tl leads to more charge transfer to the outer BaO layers. This phenomenon accounts for different charge fractions, 0.5 $\sigma_0$ for Bi-2212 *vs.* $\sigma_0$ for Tl-2212, and the marked dichotomy in $T_{C0}$. The same effect of electronegativity difference is found for the 2223 compounds $(Bi,Pb)_2Sr_2Ca_2Cu_3O_{10+\delta}$ ($\sigma = 0.5\ \sigma_0$, $T_{C0}$ = 112 K) and $Tl_2Ba_2Ca_2Cu_3O_{10}$ ($\sigma = \sigma_0$, $T_{C0}$ = 130 K).

Section 2 presents derivations based on the high-$T_C$ interlayer Coulombic pairing model, specifically comprising the new findings reported herein. These include the effect of +1 and +3 inner layer cations and their respective electronegativities relative to that of $Cu^{+2}$ in determining $\sigma$ for the compounds with mixed Tl-ion oxidation states. The model predictions are successfully tested in Section 3, wherein $\sigma$ and $T_{C0}^{Calc.}$ are self-consistently calculated for $TlBa_{1.2}La_{0.8}CuO_5$, $Tl_{0.7}LaSrCuO_5$, and $Tl_2Ba_2Ca_{\eta-1}Cu_\eta O_{2\eta+4}$ ($\eta$ = 1–3), and $f_{+1}$ = 1/3 for optimal Tl-1223 is demonstrated. The results are discussed and interpreted in Section 4 with a focus on materials issues, the significance of $\sigma_0$, and the fact that the 48 compounds shown to obey the model calculations of Section 2 form a normal distribution of $T_{C0}^{Meas.} - T_{C0}^{Calc.}$, indicating a statistically inconsequential bias in the model for $T_{C0}^{Calc.}$, as well as in measurements of $T_{C0}^{Meas.}$. Conclusions are drawn in Section 5.

## 2. Calculation models and methods

In the seminal and exemplary compound $YBa_2Cu_3O_{7-\delta}$, the interlayer Coulomb interaction occurs between the outer BaO layer in the type I BaO-CuO-BaO structure and an adjacent $CuO_2$ layer in the type II $CuO_2$-Y-$CuO_2$ structure. As studies of ultra-thin films of $Bi_2Sr_2CaCu_2O_{8+\delta}$ [23] and $YBa_2Cu_3O_{7-\delta}$ [24] have demonstrated, the minimal thickness of material sustaining superconductivity contains both reservoir structures; doped $CuO_2$ layers in isolation does not meet this requirement [25]. The doping-dependent optimization behavior, as generally exhibited, suggests that optimization is achieved at the point where the interacting charges in the two reservoirs are in equilibrium. In the high-$T_C$ cuprate compounds, being ionic solids, it is realistic to deduce that these two distinct types of charges interact via Coulombic forces. Locating this interaction on the adjacent outer layers of the two reservoir types is consistent with experiment, as shown quite dramatically for $Bi_2Sr_{1.5}Ca_{1.5}Cu_2O_{8+\delta}$ and $Bi_2Sr_{1.6}La_{0.4}CuO_{6+\delta}$, where intercalation of variously sized insulating organic molecular species between the inner type I BiO layers has virtually no effect on either $T_C$ or interaction charges, as deduced from muon-spin depolarization [26].

### *2.1 Relation of fractional charge σ to $T_{C0}$*

By adopting this model that takes the two types of charge reservoirs into account, it has been shown that the optimal transition temperature $T_{C0}$, corresponding to the highest $T_C$ for a given compound structure, is given by the algebraic equation [7],

$$T_{C0} = k_B^{-1}\,\beta\,(\sigma\eta/A)^{1/2}\,\zeta^{-1} = k_B^{-1}\,\beta\,(\ell\zeta)^{-1}\,, \tag{1}$$

where $\zeta$ is the interaction distance (measured along the transverse or hard axis), $A$ is the basal-plane area per formula unit, $\sigma/A$ is the areal charge density per type I layer per formula unit for participating charges and $\beta$ (= 0.1075 ± 0.0003 eV Å$^2$) is a universal constant, and where $e^{-2}\beta$ is approximately twice the reduced electron Compton wavelength. Superconductivity in these systems breaks down as intralayer Coulomb repulsion and charge scattering become increasingly important with decreasing $\ell$. Equation (1) has been validated and $\beta$ confirmed through extensive study of experimental results for high-$T_C$ compounds based on cuprate [7, 9], ruthenate, rutheno-cuprate [7], iron-pnictide [7], iron-chalcogenide [8], organic [7], and intercalated group-5-metal nitride-chloride [27, 28] structures, with $T_{C0}$ ranging from 6.3 K to 145 K.

A feature seemingly unique to the high-$T_C$ cuprates is their natural tendency toward a fundamental equilibrium fractional charge value equal to $\sigma_0$. This is evidenced by the fact that 10 of the 25 cuprates considered have $\sigma = \sigma_0$. Depending upon the available carriers, the structure of the reservoir layers, and the stoichiometry of the ions contained therein, the predilection toward $\sigma_0$ dictates the distribution of charge. This implies that in charge depleted systems for which $\sigma < \sigma_0$, the inner type I layers would tend to transfer some or all of their available charge to the outer layers and to the type II reservoir, so as to bring $\sigma$ as close to $\sigma_0$ as physically possible. It is also evident that in some cases charges located within the inner layer structures may not contribute to $\sigma$ at all.

Participating charges were previously introduced conceptually [7] and their presence was verified from doping behaviors in intercalated group-5-metal nitride-chlorides [27, 28]. The associated participating charge for optimal materials is determined from the difference between the dopant charge stoichiometry and the minimum stoichiometric value required for superconductivity to occur. An example is the value x = 0.163 taken relative to $x_0 = 0$ in $La_{2-x}Sr_xCuO_{4-\delta}$. This principle of charge participation, including cases of non-participation, is also extensible for considering non-optimal materials, where charge imbalance between mediating and superconducting carriers produces a sub-fraction of nonparticipating charges and is reflected in increased scattering, reduced Meissner fraction, $T_C < T_{C0}$, and associated phenomena. This interpretation was examined in the underdoped regime of the charge-compensated compound $(Ca_xLa_{1-x})(Ba_{1.75-x}La_{0.25+x})Cu_3O_y$, where $T_{C0}$ corresponds to x = 0.45 and scattering-induced pair-breaking completely accounts for the observed depression of $T_C$ for x < 0.45, as was shown in Section III C of Ref. [9].

### *2.2 Determining σ*

There are two methodologies for determining the fractional charge σ for high-$T_C$ materials, employing a first set of rules based on charge allocation and a second involving valency scaling relative to $\sigma_0$. The first rules have been employed to calculate σ for 26 of the 46 previously tested compounds: four cuprates, one ruthenate, seven Fe-pnictides, five Fe-chalcogenides, and nine intercalated group-5-metal nitride-chlorides [7-9, 27, 28]. In the present work, however, all calculations of σ for optimal cuprate compounds are calculated by scaling to $\sigma_0$ for $YBa_2Cu_3O_{6.92}$ according to the formula,

$$\sigma = \gamma\sigma_0 , \tag{2}$$

where γ is defined as the product of multiple scaling component factors $\gamma_i$ that depend upon structure and/or oxidation state according to the second set of three valency scaling rules. Since the charge doping in $YBa_2Cu_3O_{6.92}$ is largely determined by the oxygens in the CuO chains, and La-on-Ba-site substitutions are seen to suppress $T_{C0}$ [9], the rules for calculating these γ-factors have been developed with respect to the type I reservoir since the primary interest is determining the charge in the type I interacting (outer) layers.

The first valency scaling rule, denoted (2a) and given below, determines the doping charge available from cation substitution in the type I inner layer(s), where the cation acts to regulate the interchange of charge among layers. Rule (2a) originally gave the γ-factor for inner-layer cations in +3 oxidation states substituting for $Cu^{+2}$ [7]. However, single-TlO-layer cuprates may contain admixtures of $Tl^{+1}$ and $Tl^{+3}$. Accordingly, the following analysis is extended to include the γ-factor for substituting an inner-layer cation in the +1 oxidation state.

Just as the oxygen content of the CuO chain of $YBa_2Cu_3O_{7-\delta}$ determines σ, it follows logically that the same is true for a heterovalent cation substitution in the same chain layer. As determined for oxygen content, scaling provides $\sigma = \gamma\sigma_0$ in which $\gamma_1 = 1$ at $\delta_1 = 0.08$ and $\gamma_2 = 0.493$ at $\delta_2 = 0.40$, where subscripts 1 and 2 refer to the nominally 90-K and 60-K optimal-$T_{C0}$ phases, respectively [7]. The decrease $\Delta\gamma = \gamma_2 - \gamma_1 = -0.507$ originates from the increase in average valence per optimal $O^{-2}$-anion site of $\Delta v = v_2 - v_1 = (+2)(\delta_2 - \delta_1)/(1-\delta_1) = 0.696$, yielding for the $O^{-2}$ vacancies the proportionate change $\Delta\gamma/\Delta v = -0.729$. In considering a cuprate compound with similar structural characteristics as $YBa_2Cu_3O_{6.92}$, including the same inner-layer oxygen stoichiometry, an inner-layer cation mapped to the chain-Cu site having a valence greater than (or less than) the +2 valence of $Cu^{+2}$ decreases (or increases) $\sigma$, i.e., in correspondence to oxygen anion content. Thus, valency scaling provides a method for accurately determining $\sigma$ and the charge available for superconductivity in the designated cuprate.

Substituting a +3 cation at the $Cu^{+2}$ chain site has the effect of reducing $\sigma$ by the factor 1/2, since the +1 change in the valence of the inner-layer cation depletes one-half of the charge available from the native valence of +2. On the other hand, substituting a +1 cation for the $Cu^{+2}$ increases $\sigma$ by the factor 2, since a like amount of the charge available from the native valence is added. Expressing these two cases for cation substitutions as $\gamma_1 = 1/2$ for $v_1 = +3$ and $\gamma_2 = 2$ for $v_2 = +1$, one obtains the proportionate change $(\gamma_2 - \gamma_1)/(v_2 - v_1) = -0.75$, which is in good agreement with the $-0.729$ change associated with $O^{-2}$ vacancies. Thus, for the inner type I layer(s), the $\gamma$-factor for substitution of heterovalent cations is closely analogous to varying the anion stoichiometry.

Generalized to the two cases of heterovalency, the first valency scaling rule (2a) derived in the preceding analysis reads:

(2a) Heterovalent substitution in the type I inner layer(s) of a valence +3 (or +1) ion mapped to a valence +2 ion corresponding to the $YBa_2Cu_3O_{7-\delta}$ structural type introduces a factor of 1/2 (or 2) in $\gamma$.

In the case of Bi/Pb-based cuprates, the various cation substitutions in the type I inner layer also result in increasing the relative electronegativity at the associated cation site(s), decreasing further the available charge from rule (2a). Originally cast for simplicity as charge allocation expressed as an additional sharing factor of 1/2 in $\gamma$ [7], this rule (2a) modifier more accurately arises via charge transfer from the outer layers to the inner layer(s) as compensation for the factor of 1/2 charge loss, resulting in a $\gamma$-factor of (1/2)(1/2) = 1/4. Since inner-layer cation vacancies, as in the case of $Tl_{1-x}LaSrCuO_5$, tend to trap charge, they too contribute to the net transfer of carriers to the interacting outer layers, and thus would also introduce a $\gamma$-factor of 1/4 (weighted by their stoichiometric component). The ultimate physical effect of combining the increased valence and high Pauling electronegativities of Bi (2.02) and Pb (2.33) when compared to that of Cu (1.90) essentially suppresses participation of any inner-layer charges in the superconductivity, and is properly reflected by the insulating or semiconducting double BiO inner layers [29, 30], large effective mass anisotropy ratios ($m^*_c/m^*_{ab}$) of 1300 to 3000 [31, 32], micaceous crystal morphology with prominent cleavage between BiO layers, and accommodation of intercalation molecules [26]. For the Tl-based cuprates, however, the comparatively low electronegativity of Tl (1.62) is too weak to adversely affect charge transfer along the hard-axis, resulting in a comparatively small mass anisotropy (e.g., $^{205}$Tl NMR measurements reported in Ref. [33] of single crystals of $Tl_2Ba_2CaCu_2O_8$ suggest a lower effective mass ratio of ~ $10^2$), and conducting TlO layers (see, e.g., Refs. [34-36] for $Tl_2Ba_2CuO_6$, Refs. [34] and [36] in the case of $Tl_2Ba_2CaCu_2O_8$, and Ref. [36] concerning $Tl_2Ba_2Ca_2Cu_3O_{10}$). Moreover, crystals of Tl-based cuprates are non-micaceous, like $YBa_2Cu_3O_{7-\delta}$. Thus, rule (2a) without a sharing modifier fully determines the $\gamma$-factor of 1/2 associated with $Tl^{+3}$ in Tl-based superconductors.

The second valency scaling rule (2b) originates from the observation that +3 cations in the outer type I layers act to suppress superconductivity, while +2 cations tend to enhance the superconducting state. Including –2 valence anions accounts for scaling with oxygen content:

(2b) The factor $\gamma$ scales with the +2 (–2) cation (anion) structural and charge stoichiometry associated with participating charge.

Rule (2b) encompasses the full type I reservoir structure and its constituent cations, and is originally derived from the behavior of $La_{2-x}Sr_xCuO_{4-\delta}$ [37] where charge doping is observed to scale with the +2 type I outer-layer cation component. The effect for which rule (2b) applies to the outer type I layers is also seen, e.g., in $(Ca_xLa_{1-x})(Ba_{1.75-x}La_{0.25+x})Cu_3O_y$ [9] and $TlBa_{1+y}La_{1-y}CuO_5$ [10, 38], where $La^{+3}$ substituting for $Sr^{+2}$ or $Ba^{+2}$ cations acts to reduce $T_C$. The ratio of the +2 cation content compared to the full occupation of two Ba ions in $YBa_2Cu_3O_{6.92}$ gives the scaling factor. Instances of the structural aspect of rule (2b) are seen in the Bi-based cuprates, where two BiO layers replace the single CuO layer of $YBa_2Cu_3O_{6.92}$, giving a $\gamma$-factor of 2 [7]. The anion case typically applies to oxygen content.

The third valency scaling rule, which is not utilized for the present study, provides a possible way to scale between the cuprates and other high-$T_C$ families (see Ref. [7]). For compounds comprising isostructural and isovalent correspondences (of comparable electronegativities) to the type I reservoir $BaO\text{-}CuO_{0.92}\text{-}BaO$ of $YBa_2Cu_3O_{6.92}$, one has by default $\gamma = 1$.

## 3. Results

As asserted in Ref. [7], supported by experiment (see, e.g., Ref. [39]), and deduced via calculations below, the fractional charge $\sigma$ for at least five of the Tl-based cuprates is equal to $\sigma_0$. These are compounds with either a double TlO layer structure with $\eta \geq 1$, or a single TlO layer with $\eta \geq 2$. On the other hand, the compounds with the lowest $T_C$ have $\sigma < \sigma_0$ and comprise single layers of both $CuO_2$ ($\eta = 1$) and $Tl_{1-x}O$, the latter being sandwiched between two $(Ba/Sr)_{1+y}La_{1-y}O$ outer layers. A compendium of data for various cuprate and non-cuprate high-$T_C$ compounds is given in Table 1, including measured and calculated optimal transition temperatures denoted $T_{C0}^{\text{Meas.}}$ and $T_{C0}^{\text{Calc.}}$, respectively. In the following, Sections 3.1 and 3.2 present the new findings of $\sigma < \sigma_0$ in the charge depleted Tl-1201 compounds introduced in this work; Section 3.3 provides a derivation of $\sigma = \sigma_0$ for the double TlO-layer cuprates; Section 3.4 demonstrates self-consistently the approach to $f_{+1} = 1/3$ for $\sigma = \sigma_0$ at optimization in mixed oxidation state Tl-1223, and by extension Tl-1212.

### Table 1

Analysis data for 48 optimal high-$T_C$ superconducting compounds. Listed are compound formula, measured $T_{C0}^{\text{Meas.}}$, fractional charge $\sigma$ ($\sigma_0$ = 0.228), interaction distance $\zeta$, interaction-charge length $\ell$, structures of types I and II reservoirs per formula unit ($O_x$ denotes partial filling), and calculated $T_{C0}^{\text{Calc.}}$. Related compounds are grouped as containing (a) Tl, (b) $CuO_2$ (others), (c) Fe, and (d,e) other materials.

| **Superconducting compound** | $T_{C0}^{\text{Meas.}}$ **(K)** | $\sigma$ | $\zeta$ **(Å)** | $\ell$ **(Å)** | **Type I / Type II reservoir** | $T_{C0}^{\text{Calc.}}$ **(K)** |
|---|---|---|---|---|---|---|
| **(a) Thallium cuprates (Ref. [7]; this work)** | | | | | | |
| $Tl_2Ba_2CuO_6$ | 80 | $\sigma_0$ | 1.9291 | 8.0965 | BaO-TlO-TlO-BaO / $CuO_2$ | 79.86 |
| $Tl_2Ba_2CaCu_2O_8$ | 110 | $\sigma_0$ | 2.0139 | 5.7088 | BaO-TlO-TlO-BaO / $CuO_2$-Ca-$CuO_2$ | 108.50 |
| $Tl_2Ba_2Ca_2Cu_3O_{10}$ | 130 | $\sigma_0$ | 2.0559 | 4.6555 | BaO-TlO-TlO-BaO / $CuO_2$-Ca-$CuO_2$-Ca-$CuO_2$ | 130.33 |
| $TlBa_{1.2}La_{0.8}CuO_5$ | 45.4 | 0.300 $\sigma_0$ | 1.9038 | 14.684 | (Ba/La)O-TlO-(Ba/La)O / $CuO_2$ | 44.62 |
| $Tl_{0.7}LaSrCuO_5$ | 37 | 0.213 $\sigma_0$ | 1.8368 | 17.138 | (La/Sr)O-$Tl_{0.7}$O-(La/Sr)O / $CuO_2$ | 39.63 |
| $TlBa_2CaCu_2O_{7-\delta}$ | 103 | $\sigma_0$ | 2.0815 | 5.7111 | BaO-TlO-BaO / $CuO_2$-Ca-$CuO_2$ | 104.93 |
| $TlBa_2Ca_2Cu_3O_{9-\delta}$ | 133.5 | $\sigma_0$ | 2.0315 | 4.6467 | BaO-TlO-BaO / $CuO_2$-Ca-$CuO_2$-Ca-$CuO_2$ | 132.14 |
| **(b) Other cuprates, rutheno-cuprates, and ruthenates (Refs. [7, 9]; this work)** | | | | | | |
| $YBa_2Cu_3O_{6.92}$ | 93.7 | $\sigma_0$ | 2.2677 | 5.7085 | BaO-CuO-BaO / $CuO_2$-Y-$CuO_2$ | 96.36 |
| $YBa_2Cu_3O_{6.60}$ | 63 | 0.439 $\sigma_0$ | 2.2324 | 8.6271 | BaO-CuO-BaO / $CuO_2$-Y-$CuO_2$ | 64.77 |
| $LaBa_2Cu_3O_{7-\delta}$ | 97 | $\sigma_0$ | 2.1952 | 5.7983 | BaO-CuO-BaO / $CuO_2$-La-$CuO_2$ | 98.00 |
| $(Ca_{0.45}La_{0.55})(Ba_{1.30}La_{0.70})Cu_3O_y$ | 80.5 | 0.148 $\sigma_0$ | 2.1297 | 7.1176 | (Ba/La)O-CuO-(Ba/La)O / $CuO_2$-(La/Ca)-$CuO_2$ | 82.30 |
| $YBa_2Cu_4O_8$ (12 GPa) | 104 | $\sigma_0$ | 2.1658 | 5.5815 | BaO-CuO-CuO-BaO / $CuO_2$-Y-$CuO_2$ | 103.19 |
| $HgBa_2Ca_2Cu_3O_{8+\delta}$ | 135 | $\sigma_0$ | 1.9959 | 4.6525 | BaO-$HgO_x$-BaO / $CuO_2$-Ca-$CuO_2$-Ca-$CuO_2$ | 134.33 |
| $HgBa_2Ca_2Cu_3O_{8+\delta}$ (25 GPa) | 145 | $\sigma_0$ | 1.9326 | 4.4664 | BaO-$HgO_x$-BaO / $CuO_2$-Ca-$CuO_2$-Ca-$CuO_2$ | 144.51 |
| $HgBa_2CuO_{4.15}$ | 95 | $\sigma_0$+0.075 | 1.9214 | 7.0445 | BaO-$HgO_x$-BaO / $CuO_2$ | 92.16 |
| $HgBa_2CaCu_2O_{6.22}$ | 127 | $\sigma_0$+0.088 | 2.039 | 4.8616 | BaO-$HgO_x$-BaO / $CuO_2$-Ca-$CuO_2$ | 125.84 |
| $La_{1.837}Sr_{0.163}CuO_{4-\delta}$ | 38 | 0.0408 | 1.7828 | 18.673 | (La/Sr)O-(La/Sr)O / $CuO_2$ | 37.47 |
| $La_{1.8}Sr_{0.2}CaCu_2O_{6\pm\delta}$ | 58 | 0.050 | 1.7829 | 11.990 | (La/Sr)O-(La/Sr)O / $CuO_2$-Ca-$CuO_2$ | 58.35 |
| $(Sr_{0.9}La_{0.1})CuO_2$ | 43 | 0.050 | 1.7051 | 17.667 | (Sr/La) / $CuO_2$ | 41.41 |
| $(Pb_{0.5}Cu_{0.5})Sr_2(Y,Ca)Cu_2O_{7-\delta}$ | 67 | 0.375 $\sigma_0$ | 1.9967 | 9.2329 | SrO-Pb/CuO-SrO / $CuO_2$-Y/Ca-$CuO_2$ | 67.66 |
| $Bi_2Sr_2CaCu_2O_{8+\delta}$ (unannealed) | 89 | 0.500 $\sigma_0$ | 1.750 | 7.9803 | SrO-BiO-BiO-SrO / $CuO_2$-Ca-$CuO_2$ | 89.32 |
| $(Bi,Pb)_2Sr_2Ca_2Cu_3O_{10+\delta}$ | 112 | 0.500 $\sigma_0$ | 1.6872 | 6.5414 | SrO-BiO-BiO-SrO / $CuO_2$-Ca-$CuO_2$-Ca-$CuO_2$ | 113.02 |
| $Pb_2Sr_2(Y,Ca)Cu_3O_8$ | 75 | 0.500 $\sigma_0$ | 2.028 | 8.0147 | SrO-PbO-Cu-PbO-SrO / $CuO_2$-Y/Ca-$CuO_2$ | 76.74 |
| $Bi_2(Sr_{1.6}La_{0.4})CuO_{6+\delta}$ | 34 | 0.110 $\sigma_0$ | 1.488 | 24.080 | SrO-BiO-BiO-SrO / $CuO_2$ | 34.81 |
| $RuSr_2GdCu_2O_8$ | 50 | 0.250 $\sigma_0$ | 2.182 | 11.370 | SrO-$RuO_2$-SrO / $CuO_2$-Gd-$CuO_2$ | 50.28 |
| $Ba_2YRu_{0.9}Cu_{0.1}O_6$ | 35 | 0.050 | 2.0809 | 18.612 | BaO / (1/2)($YRu_{0.9}Cu_{0.1}O_4$) | 32.21 |
| **(c) Fe pnictides and chalcogenides (Refs. [7, 8])** | | | | | | |
| $La(O_{0.92-y}F_{0.08})FeAs$ | 26 | 0.020 | 1.7677 | 28.427 | (1/2)(As-2Fe-As) / (1/2)(La-2O/F-La) | 24.82 |
| $Ce(O_{0.84-y}F_{0.16})FeAs$ | 35 | 0.040 | 1.6819 | 19.924 | (1/2)(As-2Fe-As) / (1/2)(Ce-2O/F-Ce) | 37.23 |
| $Tb(O_{0.80-y}F_{0.20})FeAs$ | 45 | 0.050 | 1.5822 | 17.262 | (1/2)(As-2Fe-As) / (1/2)(Tb-2O/F-Tb) | 45.67 |
| $Sm(O_{0.65-y}F_{0.35})FeAs$ | 55 | 0.0875 | 1.667 | 13.290 | (1/2)(As-2Fe-As) / (1/2)(Sm-2O/F-Sm) | 56.31 |
| $(Sm_{0.7}Th_{0.3})OFeAs$ | 51.5 | 0.075 | 1.671 | 14.371 | (1/2)(As-2Fe-As) / (1/2)(Sm/Th-2O-Sm/Th) | 51.94 |
| $(Ba_{0.6}K_{0.4})Fe_2As_2$ | 37 | 0.050 | 1.932 | 17.482 | As-2Fe-As / (Ba/K) | 36.93 |
| $Ba(Fe_{1.84}Co_{0.16})As_2$ | 22 | 0.020 | 1.892 | 28.004 | As-2(Fe/Co)-As / Ba | 23.54 |
| $FeSe_{0.977}$ (7.5 GPa) | 36.5 | 0.023 | 1.424 | 23.883 | $Se_{0.997}$ / Fe | 36.68 |
| $Fe_{1.03}Se_{0.57}Te_{0.43}$ (2.3 GPa) | 23.3 [a] | 0.015 | 1.597 | 30.447 | $Se_{0.57}$-$Fe_{0.03}$-$Te_{0.43}$ / $Fe_{1.0}$ | 25.65 |
| $K_{0.83}Fe_{1.66}Se_2$ | 29.5 | 0.0363 | 2.0241 | 20.492 | Se-$Fe_{1.66}$-Se / $K_{0.83}$ | 30.07 |
| $Rb_{0.83}Fe_{1.70}Se_2$ | 31.5 | 0.0463 | 2.1463 | 18.289 | Se-$Fe_{1.70}$-Se / $Rb_{0.83}$ | 31.78 |
| $Cs_{0.83}Fe_{1.71}Se_2$ | 28.5 | 0.0488 | 2.3298 | 18.187 | Se-$Fe_{1.71}$-Se / $Cs_{0.83}$ | 29.44 |
| **(d) Organics (Ref. [7])** | | | | | | |
| $\kappa$–[BEDT-TTF]$_2$Cu[N(CN)$_2$]Br | 10.5 | 0.125 $\sigma_0$ | 2.4579 | 43.719 | S-chains [BEDT-TTF]$_2$ / Cu[N(CN)$_2$]Br | 11.61 |
| **(e) Intercalated group-5-metal nitride-halides (Refs. [27, 28])** | | | | | | |
| $Na_{0.16}(PC)_yTiNCl$ | 6.3 | 0.02 | 7.6735 | 25.528 | TiNCl / $Na_{0.16}(PC)_y$ | 6.37 |
| $Na_{0.16}(BC)_yTiNCl$ | 6.9 | 0.02 | 7.7803 | 25.528 | TiNCl / $Na_{0.16}(BC)_y$ | 6.28 |
| $Li_{0.08}ZrNCl$ | 15.1 | 0.00375 | 1.5817 | 54.950 | ZrNCl / $Li_{0.08}$ | 14.35 |
| $Li_{0.13}(DMF)_yZrNCl$ | 13.7 | 0.01625 | 3.4 | 26.397 | ZrNCl / $Li_{0.13}(DMF)_y$ | 13.90 |
| $Na_{0.25}HfNCl$ | 24 | 0.0125 | 1.658 | 29.864 | HfNCl / $Na_{0.25}$ | 25.19 |
| $Li_{0.2}HfNCl$ | 20 | 0.0075 | 1.595 | 38.505 | HfNCl / $Li_{0.2}$ | 20.31 |
| $Eu_{0.08}(NH_3)_yHfNCl$ | 23.6 | 0.03 | 2.6686 | 19.246 | HfNCl / $Eu_{0.08}(NH_3)_y$ | 24.29 |
| $Ca_{0.11}(NH_3)_yHfNCl$ | 23 | 0.0275 | 2.7366 | 20.113 | HfNCl / $Ca_{0.11}(NH_3)_y$ | 22.66 |
| $Li_{0.2}(NH_3)_yHfNCl$ | 22.5 | 0.025 | 2.7616 | 21.082 | HfNCl / $Li_{0.2}(NH_3)_y$ | 21.43 |

### 3.1 Charge depletion in $TlBa_{1+y}La_{1-y}CuO_5$

In the $TlBa_{1+y}La_{1-y}CuO_5$ compound, optimization occurs via partial substitution of $Ba^{+2}$ for $La^{+3}$ cations in the outer type I layers within an already suppressed charge structure, owing to the inner-layer substitution of $Tl^{+3}$ for the chain $Cu^{+2}$ of $YBa_2Cu_3O_{6.92}$. Neutron powder diffraction studies [38] and XPS core-level measurements [40] indicate that the oxidation state of Tl is near +3. Optimization for this material occurs for a nominal stoichiometry of $y \approx 0.2$ [10]. Applying the valency scaling method, σ is then determined as follows: Assuming fully occupied Tl sites and mapping the $Tl^{+3}$ ions onto the inner $Cu^{+2}O$ chain layer of $YBa_2Cu_3O_{6.92}$, rule (2a) provides a single γ-factor of 1/2 owing to the relatively low electronegativity of Tl with respect to Cu. Rule (2b) invokes another γ-factor of $(1 + y)/2 = 1.2/2$, which equals the fraction of +2 cations in the type I interaction layers relative to the two $Ba^{+2}$ ions of $YBa_2Cu_3O_{6.92}$. Combining the two γ-factors yields $\gamma = (0.5)(1.2/2) = 0.3$ and Eq. (2) gives,

$$\sigma = 0.3\,[\sigma_0] = 0.0684\,. \qquad (3)$$

Given the structural parameters [10], $A = (3.84025\ \text{Å})^2 = 14.7475\ \text{Å}^2$ and $\zeta = 1.9038$ Å, one obtains from Eq. (1) $\ell = 14.6836$ Å and a calculated $T_{C0} = 44.62$ K in excellent agreement with the measured average value of 45.4 K [10]. Considering the analysis of occupancy measurements of the nominal stoichiometry, $y = 0.18 \pm 0.05$, Eq. (3) gives calculated $T_{C0} = 44.25$ (+0.93/–0.95) K, which is in statistical agreement with the result for $y = 0.2$.

### 3.2. Charge depletion in $Tl_{1-x}LaSrCuO_5$

For the related Tl-deficient compound $Tl_{1-x}LaSrCuO_5$, optimization arises through (1) partial substitution of $La^{+3}$ for $Sr^{+2}$ with a La:Sr ratio of 1:1 and (2) the depletion of Tl ions [11], both in the type I reservoir. Although the average Tl oxidation state is predicted to lie between +1 and +3 [41], at least for full Tl occupancy, the presence of Tl vacancies and nearly full oxidation is likely to drive the oxidation state very near to +3. The optimal sample under consideration has a measured Tl occupancy of 0.64 ± 0.06 and an absence of Cu-on-Tl-site defects [11]. For the purposes of the present work, Eq. (2) is applied for the fractional Tl occupancy of 0.7, i.e., in agreement with the notation presented in Ref. [11] and the stoichiometric error quoted. Since there is only a single Sr ion compared to the two Ba ions in $YBa_2Cu_3O_{6.92}$, rule (2b) gives a γ-factor of 1/2 associated with the +2 ion content of the type I outer layers. For the inner $Tl_{0.7}O$ layer, there are two components; one associated with the occupied Tl sites and one for the vacancies. Assuming a thallium oxidation state of +3, the γ-factor for the first component from rule (2a) is (1/2)(0.7). Owing to charge trapping by inner-layer cation vacancies, and drawing the analogy with the Bi-based compounds as discussed in Section 2.2, the resulting γ-factor for the (1–x) vacancy component is (1/4)(0.3). Summing the two contributions yields $\gamma = (0.5)(0.7/2 + 0.3/4) = 0.2125$ and Eq. (2) then gives,

$$\sigma = 0.2125\,[\sigma_0] = 0.0485. \qquad (4)$$

With $A = (3.7743\ \text{Å})^2 = 14.2453\ \text{Å}^2$ and $\zeta = 1.8368$ Å for $Tl_{0.7}LaSrCuO_5$, one finds from Eq. (1) $\ell = 17.1382$ Å and a calculated $T_{C0}$ of 39.63 K, which agrees quite well with measurements showing resistivity onset at 40 K and zero at 37 K [11]. The 0.06 uncertainty in actual Tl content translates into a 1.4-K uncertainty in calculated $T_{C0}$. Note that the presence of a small $Tl^{+1}$ component would serve to elevate the calculated value. Note also that the minimum Sr content required for superconductivity is unknown and, owing to the non-insulating end material at x = 0 (i.e., $x_0$ is unknown) [11], it presently remains infeasible to apply the charge allocation methodology in calculating σ for $Tl_{0.7}LaSrCuO_5$.

### *3.3. Derivation of $\sigma = \sigma_0$ for $Tl_2Ba_2Ca_{\eta-1}Cu_\eta O_{2\eta+4}$ ($\eta$ = 1, 2, 3)*

Bond valence sum calculations of the model structure $Tl_2O_3$ suggest that the Tl oxidation state for the three optimal Tl-based compounds containing double TlO layers is very nearly +3 [42]. This finding is independently supported by experiments on $Tl_2Ba_2CuO_6$ [34, 35], $Tl_2Ba_2CaCu_2O_8$ [34, 43], and $Tl_2Ba_2Ca_2Cu_3O_{10}$ [44]; Knight shift data for all three of these compounds also show trivalent Tl [36]. Consequently, the valency scaling γ-factor for these compounds is the product of 1/2 given for +3 valence from rule (2a) and 2 from the double-layer structural stoichiometry according to rule (2b) such that,

$$\sigma = (1/2)(2)\,[\sigma_0] = \sigma_0\,, \tag{5}$$

i.e., one derives $\gamma \equiv 1$, thus proving the earlier assertion of $\sigma = \sigma_0$ for this series of Tl-based cuprates.

### *3.4. Mixed Tl oxidation states in $TlBa_2Ca_{\eta-1}Cu_\eta O_{2\eta+3}$ ($\eta$ = 2, 3)*

In the case of the two Tl-based compounds containing a single TlO layer, $TlBa_2CaCu_2O_7$ (Tl-1212) and $TlBa_2Ca_2Cu_3O_9$ (Tl-1223), the mixed-valence nature of Tl places the average Tl oxidation state between +1 and +3 [45]. As a consequence, σ becomes a function of the fractional distribution of the monovalent and trivalent Tl oxidation states, which is measured experimentally. Denoting $f_{+1}$ to be the fraction of monovalent cations in the inner TlO layer (i.e., $Tl^{+1}$), and apportioning a mixture of +1 and +3 valencies in rule (2a), one has $\gamma = (2) f_{+1} + (1/2)(1 - f_{+1})$ for these optimal compounds; from Eq. (2),

$$\sigma = \gamma\,[\sigma_0] = (1.5 f_{+1} + 0.5)\,[\sigma_0]\,. \tag{6}$$

Knowing $f_{+1}$ thus determines σ or, alternatively, $f_{+1}$ can be deduced for a given value of σ. Given the absence of +3 cations in the outer type I layers, one may assume for argument's sake that, as with their double-TlO layer counterparts in Section 3.3, charge depletion below $\sigma_0$ does not occur for these compounds; taking optimal $\sigma = \sigma_0$, Eq. (6) then reduces to $f_{+1} = 1/3$ (and $f_{+3} = 2/3$), suggesting a $[Tl^{+3}]/[Tl^{+1}]$ ratio of 2. Hence, Tl cations in mixed oxidation states are required to achieve $\gamma = 1$ optimization of the superconducting state for the Tl-1212 and Tl-1223 compounds. The following four paragraphs detail the experimental evidence corroborating a value of $f_{+1} = 1/3$, thereby establishing that $\sigma = \sigma_0$ for optimal Tl-1223 and, by extension, Tl-1212.

X-ray refinement analysis of data acquired on single crystals of $TlBa_2(Ca/Tl)_2Cu_3O_{9-\delta}$ indicate that the $[Tl^{+3}]/[Tl^{+1}]$ ratio varies with annealing in $O_2$ and $N_2$ [45, 46]. For an oxygen annealing time of 172 hours at 500 °C, $[Tl^{+3}]/[Tl^{+1}]$ was found to be 0.792/0.208(10), corresponding to $f_{+1} = 0.208 \pm 0.010$, and $T_C$ = 116.3 K, where the uncertainty in $f_{+1}$ is given by the constrained ratio. Subsequent to annealing in nitrogen for 224 hours at the same temperature, $T_C$ increased to 120.6 K (lower than $T_{C0}$ = 133.5 K, see Table 1), with $[Tl^{+3}]/[Tl^{+1}]$ equal to 0.762/0.238(13), or $f_{+1} = 0.238 \pm 0.013$. These results were derived from constrained fits where the $Tl^{+1}$ cations are assumed more likely to occupy the four-fold symmetry sites [45, 46]. Although the tested samples are non-optimal Tl-1223 materials, owing to significant Tl-on-Ca site defects (~5% from Refs. [45] and [46]) and $T_C < T_{C0}$, the measurements point to a trend of $f_{+1}$ increasing with increasing $T_C$.

Research on a sample of Cu-substituted (Tl/Cu)-1223 reported a measured (diamagnetic onset) $T_C \approx 131$ K after annealing in nitrogen [47]. The as-synthesized material with a $T_C$ of 97 K is oxygen overdoped. From composition analysis, two stoichiometric possibilities were proposed: $(Tl_{0.79}Cu_{0.21})Ba_2Ca_{1.84}Cu_{3.16}O_{9\pm\delta}$, which suggests that a portion of the Cu compensates for deficiency in Tl, and $(Tl_{0.63}Cu_{0.37})Ba_2(Ca_{1.84}Tl_{0.16})Cu_3O_{9\pm\delta}$, where the excess Cu occupies the Tl sites and the remaining Tl fills the Ca vacancies. Although the authors of Ref. [47] favor the former stoichiometry, claiming an absence of conclusive evidence of Tl at the Ca sites in this system, the problem of Tl substituting for Ca is well known in the Tl-based compounds; since estimates place a substitution rate of at least 5% for Tl-1223 and up to 28% in the double-TlO-layer compounds [45, 46], it is reasonable to expect similar defects manifesting in the Cu-substituted compounds. Thus, the latter stoichiometry with all of the Cu in excess of 3.00 substituting for Tl in the type-I (Tl/Cu)O inner layer, and 0.16 of the remaining Tl excess occupying Ca sites, is treated here; the alternative with an unsatisfied Ca deficiency and an unassigned Cu excess is considered unrealistic.

From XPS measurements of the Tl-4f$_{7/2}$ core-level binding energies in (Tl/Cu)-1223, the fraction of monovalent thallium ($Tl^{+1}$) increases from about 0.08 in the as-synthesized material to about 0.50 after the 550-°C reduction anneal (from Table II of Ref. [47] and an XPS peak midway between $Tl^{+1}$ and $Tl^{+3}$ [48]). With Cu substituting for Tl in the inner type I layer, and taking the average Cu oxidation state to be at least +2 in the (Tl/Cu)O layers, as inferred from the $Cu^{+2}O$-like Cu $2p_{3/2}$ XPS spectra [48], the fraction of monovalent cations in the type I inner layer, $f_{+1}$, is determined as the product of the type-I Tl stoichiometry of 0.63 and the $Tl^{+1}$ fraction; the results are 0.053 and 0.315 for the as-synthesized and nitrogen-annealed materials, respectively. An uncertainty in $f_{+1}$ for both materials of ±0.033 is estimated from composition analysis of similarly prepared materials (averaging uncertainties for $T_C \geq 130$ K compositions in Table 2 of Ref. [49]); the uncertainty in $Tl^{+1}$ fraction is estimated from the uncertainty in Tl content [47]. It is also clear that the variable $T_C$ observed in (Tl/Cu)-1223 is decoupled from charges on the $CuO_2$ layers, since no binding energy shifts occur in the Cu $2p_{3/2}3d^9$ XPS peaks [48], in contrast to the systematic shifts of these peaks with oxygen doping in $YBa_2Cu_3O_{7-\delta}$ [50].

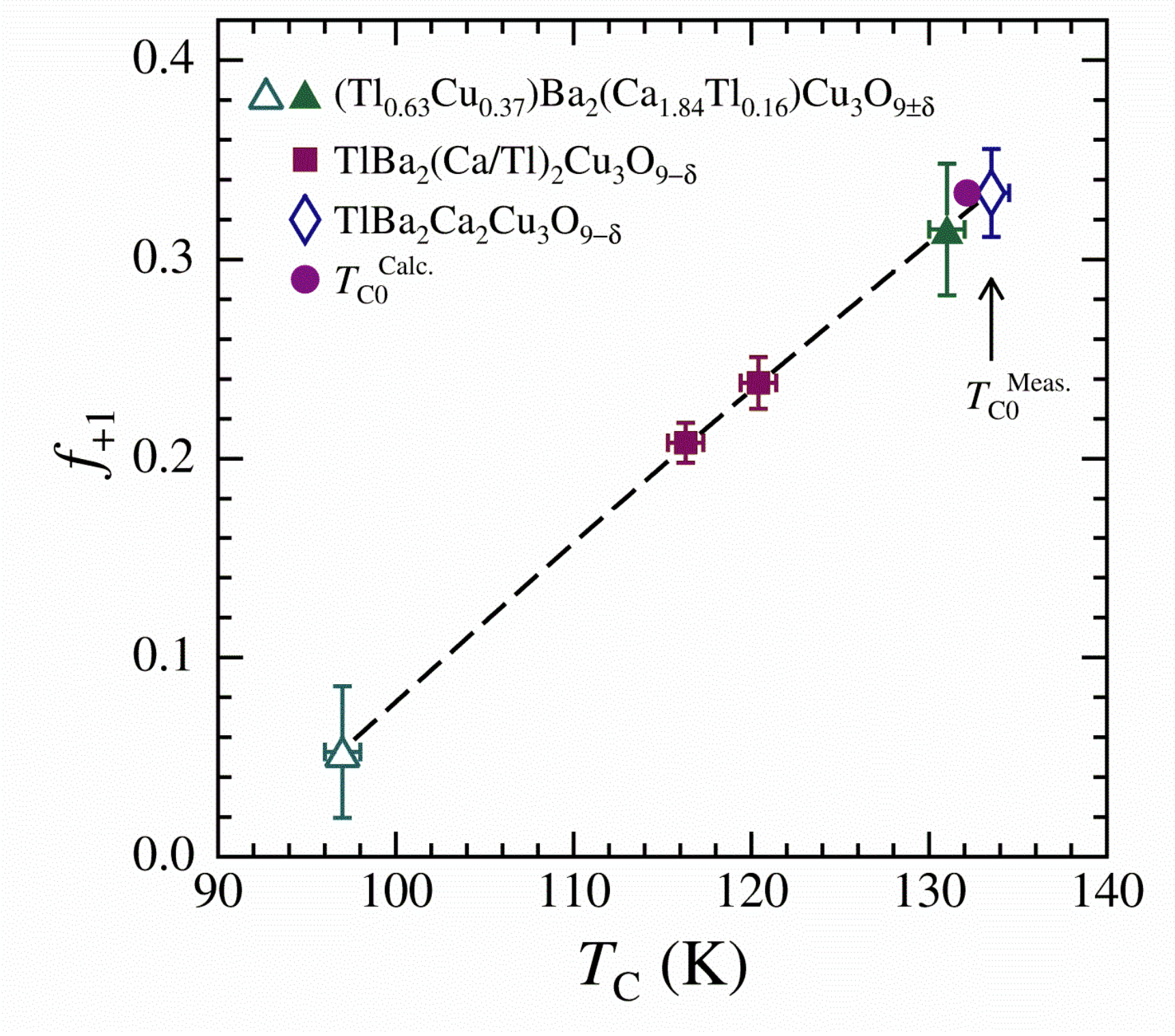

**Fig. 1.** Plot of $Tl^{+1}$ fraction $f_{+1}$ versus measured $T_C$ for as-synthesized and annealed $(Tl_{1-x}Cu_x)Ba_2(Ca/Tl)_2Cu_3O_{9\pm\delta}$ (open and filled triangles, respectively) [47], $TlBa_2(Ca/Tl)_2Cu_3O_{9-\delta}$ (filled squares) [45, 46], and $TlBa_2Ca_2Cu_3O_{9-\delta}$ (open diamond) [49]; the dashed line through the data represents the trend. The filled circle denotes calculated $T_{C0}^{Calc.}$ for $f_{+1}$ = 1/3 predicted for $TlBa_2Ca_2Cu_3O_{9-\delta}$ with measured $T_{C0}^{Meas.}$ indicated.

Experimental $T_C$ and $f_{+1}$ data for as-synthesized and annealed $(Tl_{0.63}Cu_{0.37})Ba_2(Ca_{1.84}Tl_{0.16})Cu_3O_{9\pm\delta}$, along with the two non-optimal Tl-1223 samples, are compared in Fig. 1 to optimal $TlBa_2Ca_2Cu_3O_{9-\delta}$ of measured composition $Tl_{1.05(3)}Ba_{1.99(1)}Ca_{1.96(2)}Cu_{3.00(6)}O_{9-\delta}$, where the 0.05(3) excess Tl is presumed to occupy 0.04(2) Ca sites [49]. The latter sample was cited in the original work as being representative of the optimal state with $T_{C0} = 133.5$ K [7]. Although no XPS data are shown or cited specific to this sample, Ref. [49] states that XPS shows the average Tl oxidation state approaching from +3 to +2, which we interpret as approaching +2 from above and assume it to be near the optimal value of +2 1/3 calculated for $f_{+1} = 1/3$. The uncertainty in $f_{+1}$, derived from the uncertainties in cation ratios, is found to be ±0.022, and shown as the error bars for the open-diamond symbol with indicated $T_{C0}^{\mathrm{Meas.}}$ in Fig. 1. The estimation $T_{C0}^{\mathrm{Calc.}} = 132.14$ K (filled circle) is also included for $f_{+1} = 1/3$ (using structural refinement data from a different Tl-1223 sample) [7], along with a trend line through the data, lending strong support for an optimal value of $f_{+1} = 1/3$ and validating $\sigma = \sigma_0$ for optimal $TlBa_2Ca_2Cu_3O_{9-\delta}$.

The data of Fig. 1 also suggest $T_C \leq 90 - 91$ K as $f_{+1} \rightarrow 0$, possibly corresponding to a limiting non-optimal, overdoped material with depressed $T_C$. Considering a hypothetical optimal Tl-1223 sample with full $Tl^{+3}$ occupancy in the inner type I layer, Eq. (6) reduces to $\sigma = 0.5\ \sigma_0$ and the associated result $T_{C0}^{\mathrm{Calc.}} = 93.4$ K.

## 4. Discussion

The results presented above confirm the important roles played by elemental electronegativity and the cation oxidation state in governing the interaction charge density available for superconductivity. This more complete understanding explains the origin of the earlier assertation that $\sigma = \sigma_0$ for many of the Tl-based cuprates, particularly for five compounds comprising multiple TlO and/or $CuO_2$ layers [7]. The Tl-based compounds are unique among the various cuprate groups by offering not only the widest ranges in $T_{C0}$ (a maximum-to-minimum ratio in $T_{C0}$ of $r_{Tc0} = 3.5$) but also of $\sigma$ (corresponding ratio $r_\sigma = 4.7$), which are well suited for examining trends in the interaction charge. The other cuprate groups of compounds in Table 1 also have appreciable values of $r_{Tc0}$ and $r_\sigma$, being 3.3 and 4.6 (Pb- and/or Bi-based), 1.7 and 2.3 (Y- or La-based with CuO chains), 1.5 and 1.4 (Hg-based), and 1.5 and 1.2 (La/Sr-based), respectively. Comparing these five cuprate groups, $r_{Tc0}$ tends to increase with $r_\sigma$, reflecting the form of Eq. (1).

### *4.1. Accuracy in calculated $T_{C0}$*

Data for the $N = 48$ compounds given in Table 1 provide a sample of sufficient size for conducting a detailed examination of the statistical distribution in the differences between measured and calculated $T_{C0}$. Sorting results for ($T_{C0}^{Meas.} - T_{C0}^{Calc.}$) in monotonically ascending sequence, one obtains a deviation $(\Delta T_{C0})_k \equiv (T_{C0}^{Meas.} - T_{C0}^{Calc.})_k$ for the difference between measured and calculated $T_{C0}$ for a given compound $k$, where $k = 1, \ldots N$, and accordingly defines a cumulative number $N_k = k$. The distribution of $N_k$ versus $(\Delta T_{C0})_k$ thereby obtained is shown in Fig. 2 and corresponds to root-mean-square deviation $\langle \Delta T_{C0} \rangle_{rms}$ = 1.29 K. The seven Tl-based cuprate compounds are designated by individual symbols as indicated in the figure legend; the 18 other cuprate compounds are indicated by open-circle symbols and the remaining non-cuprate compounds by open-triangle symbols. For a normal distribution of errors, the number $N_k$ is expected to approach the integral of a Gaussian function and may be modeled as $N_k = (N/2)$ $[\mathrm{erf}(t_k) + 1] + N_0$, where erf is the error function of argument $t_k = 2^{-1/2}\delta_{Tc}^{-1}(\Delta T_{C0})_k + t_0$, and where $\delta_{Tc}$ is the standard deviation or width of the distribution and $t_0$ is an offset; $N_0 \sim \pm 1$ is the constant of integration. Fitting this model function for $N_k$ to the data yields the results $\delta_{Tc}$ = 1.31(2) K, $t_0$ = 0.15(2), and $N_0$ = 0.3(4). The fitted model function is represented by the continuous solid curve in Fig. 2; the rms deviation between the curve and data points in Fig. 2 is 1.1 in cumulative number. One observes that the data points closely follow the model function, as suggested by the rms deviation near unity and quantified in finding the fitted $\delta_{Tc}$ to be statistically equivalent to $\langle \Delta T_{C0} \rangle_{rms}$. The fitted constant of integration $N_0$ is small and accounts for sparse data in the low-probability tails of the distribution. The parameter $t_0$ corresponds to a temperature-scale offset of $2^{1/2}t_0\delta_{Tc}$ = 0.28(4) K, which, being well within one standard deviation ($\delta_{Tc}$ and $\langle \Delta T_{C0} \rangle_{rms}$), indicates statistically inconsequential bias in the model for $T_{C0}^{Calc.}$, as well as in measurements of $T_{C0}^{Meas.}$.

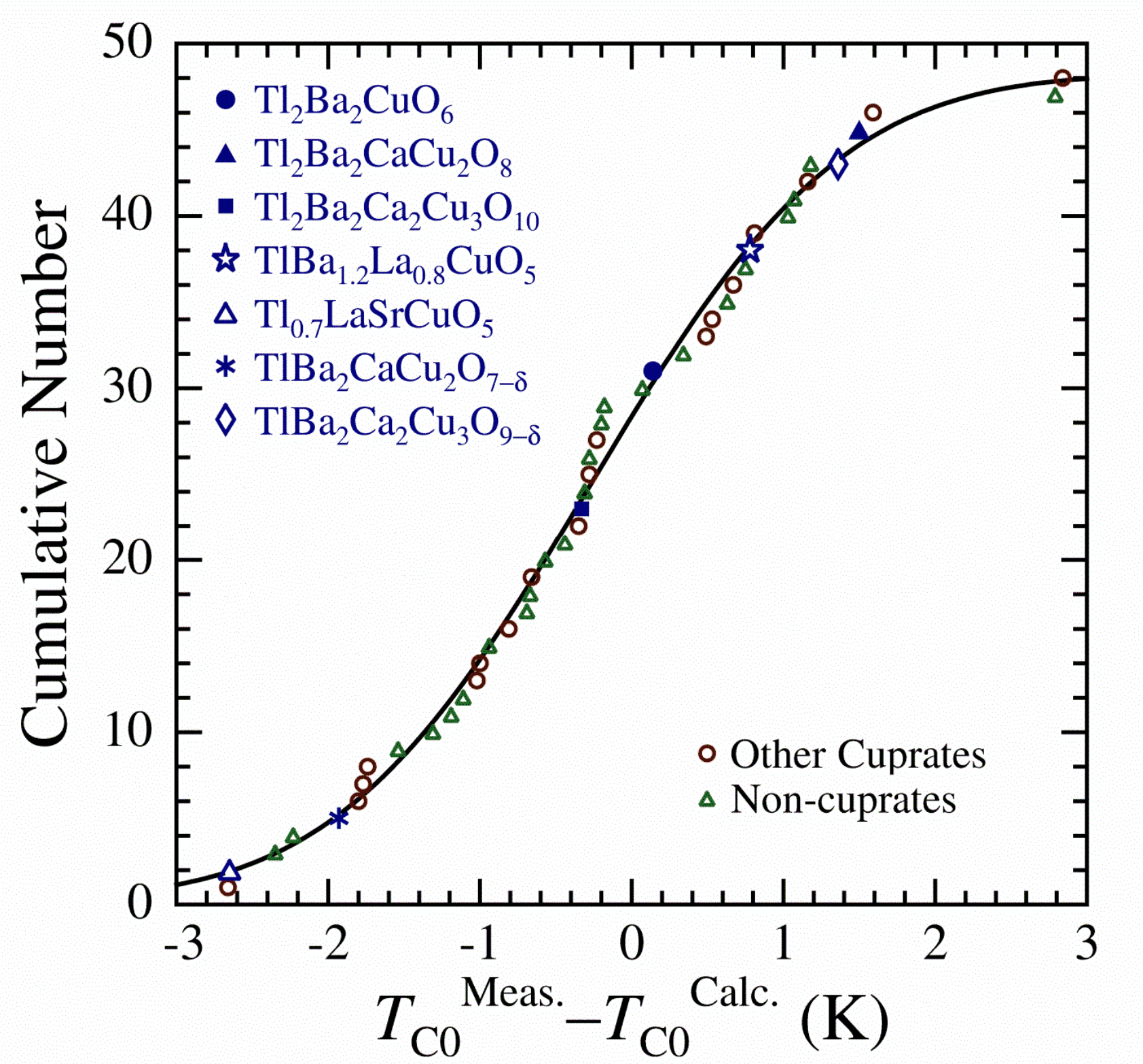

**Fig. 2.** Cumulative number distribution for the differences between measured $T_{C0}^{Meas.}$ and calculated $T_{C0}^{Calc.}$ for the 48 compounds listed in Table 1 and represented by symbols indicated in legend. Curve is the modeled error function distribution of fitted width 1.31(2) K, as discussed in the text.

### *4.2. Electronegativity and the charge-regulating type I cations*

In earlier work [19-22], the average electronegativities of the cations ($\chi_c$) and/or anions ($\chi_a$) of various high-$T_C$ compounds, calculated using several different methods (arithmetic or geometric averaging, ionization potentials, etc.), were considered and compared to corresponding transition temperatures. As these studies did not include the two reservoir-type nature of these materials and did not discriminate with regard to charge reservoir or microstructure therein, it is not surprising that they enjoyed only limited success. It is interesting to note, however, that the average $\chi$ values for most of the cuprates studied are greater than 2, falling outside the range (1.3–1.9 [20]) of the superconducting elements in a manner consistent with the cuprates being comparatively poorer metals [19, 20]. This result is basically confirmed in Ref. [21]. Tendencies for $T_C$ to be increasing with ($\chi_a - \chi_c$) and [$\chi_c/(\chi_c + \chi_a)$] above threshold levels of 1.90 and 0.176, respectively, were also indicated [22, 51].

The present work recognizes the unique two-reservoir character inherent in all high-$T_C$ superconductors, and the complimentary roles played by the individual ions and layers of the two charge reservoirs. Alternative treatments considering only $CuO_2$ layers are by comparison incomplete. Supplemental to bond-valence sum (BVS) calculations [7], electronegativity considerations can provide fruitful insight. By combining electronegativities of the elements constituting the cuprate compounds in Table 1, the type I reservoirs are found to be electropositive relative to the type II reservoirs, showing that the type I reservoir is cation-like and the type II reservoir is anion-like. Considering just the adjacent interacting layers, the interacting layer in the type I reservoir is electropositive relative to the (outer) $CuO_2$ layer by one or more Pauling units, owing to Ba, Sr, La, Nd, and Ce all having smaller electronegativities than Cu. Hence, the cuprates have positive or hole-like interaction charges in the type I outer layers with negative or electron-like interaction charges residing in the (outer) $CuO_2$ layers.

Serving as the doping source in the high-$T_C$ cuprates, the type I reservoir regulates the charge density of the interaction layers. Depending upon the relative elemental electronegativities and oxidation states of the type I inner-layer cations, the carriers available for superconductivity may be supplemented or depleted. This effect is rather dramatically displayed by comparing the structural, stoichiometric, and electronic properties of $TlBa_{1.2}La_{0.8}CuO_5$ [10], $Tl_{0.7}LaSrCuO_5$ [11], and $La_{1.837}Sr_{0.163}CuO_{4-\delta}$ [37]. While for these compounds the values of $T_{C0}$, $\zeta$, and $\ell$ are comparable (Table 1), an inner TlO layer regulates doping in the Tl-based compounds, whereas (La/Sr)O layers regulate doping in $La_{2-x}Sr_xCuO_{4-\delta}$. Here, the (La/Sr)O layers are regarded as two outer type I layers in the absence of an inner layer. From the perspective of the model of Section 2, the insertion of a $Tl^{+3}O$ inner layer between a pair of (La/Sr)O or (La/Ba)O outer layers drives the material into the underdoped regime, leaving the type II reservoir with excess charge relative to the type I reservoir. This is clear, given that $TlBa_{1+y}La_{1-y}CuO_5$ is essentially insulating for all $y \leq 0$. In order to recover the optimal superconducting state, the charge distribution is then rebalanced by a compensating increase in the +2 cation content in the type I outer layers.

Since $La_{2-x}Sr_xCuO_{4-\delta}$ is insulating at $x_0 = 0$, all of the carriers available for superconductivity, both holes and electrons, must originate from Sr doping, with an optimal value of $x - x_0 = 0.163$. On the other hand, for $TlBa_{1+y}La_{1-y}CuO_5$, which can be equivalently written as $TlLa_{2-x'}Ba_{x'}CuO_5$ with $x' = 1 + y$, the presence of the intervening $Tl^{+3}O$ layer and the compensating increase in the type I outer-layer +2 cation content relative to $La_{1.837}Sr_{0.163}CuO_{4-\delta}$ transforms the insulating-phase baseline from $x_0 = 0$ (i.e., absence of an alkaline-earth in $La_2CuO_4$) to $x' = 2$ or $y_0 = 1$ (i.e., absence of a lanthanide in $TlBa_2CuO_5$), so that optimization occurs at $(y - y_0) = 0.2$. Similarly, the $Tl_{0.7}LaSrCuO_5$ compound optimizes at a Sr content equal to that of La, well above the x = 0.163 level of $La_{1.837}Sr_{0.163}CuO_{4-\delta}$, but with a depleted Tl site occupation (obviously, some of the charge involved in superconductivity is regulated through this depletion). While the large Sr level at optimization suggests a correspondingly large minimal Sr content necessary for superconductivity, the actual value is unknown (see Section 3.2).

The comparatively low elemental electronegativity of Tl relative to Cu also has important implications for the other Tl-based cuprates: For the compounds with double TlO layers, the Tl oxidation state is typically +3, such that application of rules (2a) and (2b) gives $\gamma = (1/2)(2) = 1$; from Eq. (2) one then obtains $\sigma = \gamma\sigma_0 = \sigma_0$ and derives the original ansatz [7]. In the case of the single-layer Tl compounds with $\eta > 1$, the oxidation states of the Tl ions are predicted to be distributed approximately equally between +1 and +3, with a net $\gamma$-factor of unity, again resulting in $\sigma = \sigma_0$. Consequently, the principle reason the $T_{C0}$ value for $Tl_2Ba_2CaCu_2O_8$ is greater than that of $YBa_2Cu_3O_{6.92}$ is structural, particularly in regard to their values of $\zeta$ (2.0139 versus 2.2677 Å, respectively).

The data for the Hg-based cuprates in Table 1 show $\sigma$ given as $\sigma_0$ plus an additive term accounting for excess oxygen accommodated in the $\eta$ = 1,2 compounds; the $\sigma_0$ value here follows from the single $HgO_x$ layer with partial O filling corresponding to the $CuO_{0.92}$ chain layer in $YBa_2Cu_3O_{6.92}$ and the oxidation state of Hg limited to +2 [7].

### *4.3. Value of $\sigma_0$ and quantum critical points*

Theoretically constructed phase diagrams for temperature *vs*. doping in the cuprate superconductors have associated the region of optimal doping with a quantum critical point at zero temperature [12, 52-55]. A dome-shaped curve describing the variation of $T_C$ with $CuO_2$ doping encloses the region where the quantum criticality is expected to occur and, in the vicinity of the optimal doping level, strange-metal linear-in-$T$ resistivity characterizes the normal state. Pseudo-gap phenomena and indications of phases competing with superconductivity are observed for underdoping, while overdoping induces a crossover to a conventional Fermi-liquid metal [12, 55]. Experimentally, the variation of $T_C$ with doping is approximate, owing to inhomogeneities and indistinct $T_C$ observed for materials with $T_C < T_{C0}$, and dependence on $CuO_2$ doping evidently does not follow a universal function [13]. Overdoped $Tl_2Ba_2CuO_{6+\delta}$ presents a particularly interesting case for theoretical consideration, because it exhibits superconductivity coexisting with a complete Fermi surface [13], yet also a linear-$T$ term in the resistivity and temperature-dependent Hall coefficient at low temperature [56].

Properties of the cuprate layers in the type II reservoir have been widely studied. Of particular interest here are theoretical demonstrations that ordering and electronic fluctuations are capable of mediating high-$T_C$ superconductivity; it has been put forth that certain modes of these fluctuations were already observed experimentally [15]. Most importantly, experiments sensitive to the whereabouts of the doping charges, such as by O-1s and Cu-2p near-edge x-ray absorption (XAS) fine structure, find that superconductivity is not induced by solely charging cuprate layers; instead, charging needs to be produced in the outer interacting layers of the type I reservoir (e.g., in Ba-O layers containing Ba and apical oxygen) [57]. These XAS results, together with the hole-like character of the type I reservoir discussed above, signify a superconducting condensate sustained by the type I reservoir and Coulombic mediation involving the electron-like type II reservoir. The presence of hole-like carriers coexisting with more strongly scattering electron-like carriers accounts for temperature dependences observed in the Hall coefficient ($YBa_2Cu_3O_{7-\delta}$ and $Bi_2Sr_2CaCu_2O_{8+\delta}$ [58]; $TlBa_{1+y}La_{1-y}CuO_5$ [10]; $Tl_2Ba_2CuO_{6+\delta}$ [56] and $TlSr_2CaCu_2O_{7+\delta}$ [59, 60]) and its anisotropy in sign [58]. Thus the charge doping in the outer layers of the type I reservoir, rather than that of the cuprate layers, is more appropriate for modeling temperature-doping phase diagrams.

That the optimal charge is $\sigma_0$ for a significant number of diverse cuprates suggests $\sigma_0$ may be connected to the $T$ = 0 quantum critical point considered in theory [12, 15, 53, 55]. Results of XAS experiments on $YBa_2Cu_3O_{6.91}$ (e.g., see Table 1 of Ref. [57]; similar, though scaled, results are in Table II of Ref. [61]) have determined an average charge of 0.236 ± 0.006 (per layer per formula unit containing two $CuO_2$ layers, two BaO layers, and one CuO chain layer), where the 0.03 error quoted in the Table 1 caption of Ref. [57] is assumed to be divided equally among the five layers; this experimentally derived result differs by only one standard deviation from the value of $\sigma_0$ = 0.228 for optimal $YBa_2Cu_3O_{6.92}$, obtained by allocating the O-doping charge equally among the five O-containing layers [7].

These experimental findings lend further support to the correctness of the model in Section 2 in regard to identifying the spatially separated interacting charges, relating doping to the superconducting interaction charge, and explaining optimization. A minimalist view of the superconductivity as involving only one type of carrier and location runs counter to this and previously reviewed experimental evidence [7].

### *4.4. Materials optimization*

Uncertainties in the Ba/La ratio in $TlBa_{1+x}La_{1-x}CuO_5$ significantly affect the net doping calculated. The maximum in $T_C$ versus y exhibited in Fig. 3 of Ref. [10], which includes data from earlier work [62], is certainly sufficiently broad to support an optimal y-value between 0.2 and 0.3. Indeed, the anomalous screening onset at about 49.5 K in the magnetic susceptibility for y = 0.2 in Fig. 2 of Ref. [10] and the highest $T_C$ (~40 K) shown in Fig. 4 of Ref. [40] for a nominal Ba content of 1.3 may be indicative of this possibility. There may also exist Tl vacancies which, if taken into account, would affect the value of σ. The authors of Ref. [38] estimate a ±2% error on the Tl occupancy away from unity, while those of Ref. [10] state an uncertainty of ±0.05. Such small defect levels, however, would present comparatively small corrections to the result σ = 0.0684 of Eq. (3).

Other results suggest possible synthesis difficulties as well. For example, while samples of $TlBa_{2-x}La_xCuO_{5-\delta}$ show the highest resistively determined $T_C$ (=42 K) occurring for x = 0.6 [63], the broadened resistive and magnetic transitions clearly indicate an inhomogeneous superconducting state. Further, a superconducting onset at 52 K is claimed for $TlBa_{1.2}La_{0.8}CuO_5$, based on measurements of flux expulsion [38]; unfortunately the data are not shown, preventing a determination of material quality and, e.g., the phase purity of the specimen.

Superconductive properties and limited structural information (absent refinement data) of a related compound with nominal stoichiometry $TlBaSrCuO_{5-\delta}$ have also been reported [64]. The few available results were obtained for inhomogeneous and non-optimal samples, as indicated by a resistance transition broadened by about 30%, thereby precluding a useful determination of an optimal σ. Also, compositions of the form $Tl_{1-x}LaSrCu_{1+x}O_5$ containing substitution of Cu for Tl were studied for x from 0.0 to 0.6 [11]. When compared to $Tl_{0.7}LaSrCuO_5$, these materials exhibit reduced and broadened superconducting transitions, together with appreciably higher normal-state resistivities that increase with x. Thus, these materials are presently regarded as containing Cu-on-Tl-site defects and non-optimal superconductivity.

An unrelated, but interesting, materials-related problem also occurs in $Bi_2Sr_2(Ca,Y)_xCu_2O_{8+\delta}$, where excess oxygen located in the $Bi_2O_2$ block [5], as well as cation stoichiometry and substitutional defects [65], are correlated with a strong incommensurate modulation along the *b*-axis of the basal plane. This extra oxygen content, which appears to be enhanced (along with $T_C$) by increasing the Ca vacancies (for x < 1, with zero Y content) or the Y/Ca substitution ratio (x ≈ 1) (see, e.g., Refs. [66] and [5]), introduces additional carriers for which the cation scaling rules in Section 2.2 alone do not account (accurate knowledge of the excess oxygen content in the $Bi_2O_2$ block would be required in order to apply anion scaling). Variation of $\zeta$ for non-stoichiometric compositions would also come into play. Consequently, cation scaling to $YBa_2Cu_3O_{6.92}$ according to Section 2.2 is applied to only the near-stoichiometric $Bi_2Sr_2CaCu_2O_{8+\delta}$ compound (as-grown and unannealed), assuming minimal extra-oxygen enhancement in carrier density, and having a measured transition of 89 K (see, e.g., Ref. [5, 65]). Utilizing more recent structural refinement data specific to a near-stoichiometric single-crystal sample at 12 K [65], Eq. (1) gives $T_{C0}{}^{Calc.}$ = 89.32 K (the value of 86.65 K originally published in Ref. [7] was obtained using structural data for a non-stoichiometric powder sample). The new parameters are listed in Table 1.

### *4.5. Implication for new materials*

Equation (1) contains the key essentials for investigating new high-$T_C$ materials. In considering methods for discovering and synthesizing new superconductors, an actionable objective for reaching higher values of $T_{C0}$ is found by striving for smaller values of the $\ell$-$\zeta$ product. Although the practical limitations on the lowest attainable values of $\ell\zeta$ are presently unknown, owing to unspecified composition and structure, one could nevertheless invoke a conjecture based on data for the known optimal superconductors in Table 1 by identifying a set of extremal values for the parameters that determine $\ell$ and $\zeta$. Considering that $\ell = (\sigma\eta/A)^{-1/2}$, this approach selects the largest $\sigma$ = 0.316 from the 2212 Hg-cuprate, the largest $\eta$ = 3 among optimal cuprates, and the smallest $A$ = 13.119 Å$^2$ and $\zeta$ = 1.424 Å from $FeSe_{0.977}$ [8]. This yields $\ell$ = 3.72 Å, which satisfies the limitation $\ell > \zeta$ [7], and $\ell\zeta$ = 5.297 Å$^2$. From Eq. (1) an estimated 235 K for optimal $T_{C0}$ is thereby deemed feasible. A lower limit for $\ell\zeta$ was previously considered by taking $\ell = \langle\zeta\rangle$ from the average $\zeta$ among known optimal compounds [7]; from Table 1 one has $\langle\zeta\rangle$ = 1.933Å, $\ell\zeta$ = 3.736 Å$^2$, and an upper limit of 334 K for $T_{C0}$. Approaching this upper limit entails synthesizing a material with an optimal fractional charge $\sigma$ significantly greater $\sigma_0$ as well as smaller values of the basal-plane area $A$ and the interaction distance $\zeta$.

Numerous authors have considered some already investigated materials as containing inclusions of superconducting phases with values of $T_C$ significantly higher than a known optimal $T_{C0}$ [67, 68]. Although such a view is consistent with the feasibility estimates discussed above, the potentially superconducting phases are actually unidentified and probably exist in trace quantities. It may be reiterated that the experimental data compiled for Table 1 is derived solely from the best available single-phase bulk compounds. In hypothesizing the structure and doping of a new superconducting phase, the optimal value of $\sigma$ is constrained by balanced charge reservoirs concomitant with weak pair-breaking. Owing to fluctuations and disorder in mixed phase materials, transition temperatures inferred from various experiments are likely to be lower than hypothesized. One can specify, as discussed below, the minimum amount of material required for observing superconductivity by considering the properties of thin superconducting crystals in the context of the model of Section 2.

In an earlier study, a theoretical sheet transition temperature $T_{CS}$ ($< T_{C0}$) was determined for a crystal comprising the layers in a single formula unit, which constitutes a thickness defined as the periodicity distance $d$ [25]. From the microscopic $T_{C0}$ of Eq. (1) and theory for phase transitions in two-dimensional superconductors [69], theoretical values $T_{CS} = \alpha T_{C0}$ with $\alpha$ varying from 0.51 to 0.95 are obtained for the optimal superconductors in Table 1. Experiments on several thin crystal structures of thickness $d_F$ approaching the periodicity $d$ were shown to be consistent with $T_{CS}$ and the presence of disorder. Careful analyses of these thin film studies find that the onset of superconductivity occurs at a threshold or minimum thickness $d_F \approx d$. Recognizing that $d$ contains the combined thickness of the two reservoirs, these observations demonstrate that one formula unit containing both superconducting and mediating reservoirs is the prerequisite for superconductivity to occur. Therefore, the required presence of at least one set of adjacent types I and II charge reservoirs must be taken into account when identifying and designing new high-$T_C$ materials.

The importance of electronegativity and cation oxidation state in determining the charge available for the high-$T_C$ superconducting state suggests that higher transition temperatures may be achieved by developing a new material using cations of elements with electronegativities smaller than those in typical cuprates. Since electronegativity increases in magnitude from the lower left to the upper right of the periodic table of the elements, one would naturally gravitate toward low-electronegativity elements as good candidates. Acting on this premise and realizing that any model high-$T_C$ structure must contain two physically-separated charge reservoirs, multiple-layer (e.g., $\eta > 1$) perovskite-like structures have been designed with predicted transition temperatures above 145 K [70].

The Coulombic pairing model presently being considered for new materials involves a local interaction, requiring only an interface comprising two areal charge densities of opposite sign, separated by a distance $\zeta$ that does not require extended 2D sheets of charge; the mediating charges may even occupy localized states as long as they retain the ability to exchange momenta. It is therefore reasonable to assume that this model may be applied to the alkali-metal doped Buckminsterfullerenes (see, e.g., Ref. [71]). These 3D macromolecular compounds exhibit several key characteristics of high-$T_C$ superconductors and present clear geometrical possibilities for 2D interfacial Coulomb interactions.

## 5. Conclusions

By possessing the broadest ranges in optimal transition temperature $T_{C0}$ and fractional charge $\sigma$, the Tl-based cuprates are uniquely suited for studying relationships between fundamental chemistry in the layered crystal structures and the formation of high-$T_C$ superconductor states. As shown to be general for the high-$T_C$ cuprates, the electronegativity and oxidation state of the element occupying an inner-layer cation site in the type I charge reservoir determines the fractional charge and therefrom the available superconducting interaction charge density $\ell^{-2}$. Compounds containing a type I inner layer element that is electropositive relative to Cu, such as Tl, exhibit greater efficacy in charge transfer between layers and the types I and II charge reservoirs, thereby inducing larger values of $\sigma$ (and $T_{C0}$), when compared to more electronegative elements such as Bi and Pb. The low electronegativity of Tl (compared, e.g., to Bi and Pb) with respect to that of Cu explains the disparate $T_{C0}$ values observed between the Bi/Pb-Based and Tl-based compounds and is shown to have other important implications for the Tl-based cuprates. The enhancement in $\sigma$ associated with $Tl^{+1}$ (and analogous inner-layer cations) is also found to be uniquely essential in explaining the behavior of mixed-valence compounds.

The two Tl-1201 compounds introduced in this investigation, $TlBa_{1.2}La_{0.8}CuO_5$ and $Tl_{0.7}LaSrCuO_5$, both illustrate the effects of Tl-ion oxidation states near +3 (as compared to $Cu^{+2}$) and the low electronegativity of Tl (1.62 *vs.* 1.9 for Cu). Charge depletion in these compounds reduces σ relative to $\sigma_0$ by the γ-factors of 0.3 and 0.2125, respectively, leading to their relatively low values for $T_{C0}$ (< 50 K). Hole trapping at Tl vacancies in the $Tl_{0.7}LaSrCuO_5$ compound is treated by likening the vacancies to high-electronegative heterovalent (specifically, +3) substitutions with respect to the application of valence scaling rule (2a).

For the compounds with double TlO layers, the Tl valence is nominally +3. Application of valence scaling rules (2a) and (2b) provides compensating γ-factor components of 1/2 and 2, respectively, to give $\gamma = (1/2)(2) = 1$ and $\sigma = \gamma\sigma_0 \equiv \sigma_0 = 0.228$, as previously asserted in Ref. [7], but originally left unproven. In the case of the single-Tl-layer compounds with $\eta > 1$, the oxidation states of the Tl ions are distributed between +1 and +3. Given a value of $\sigma = \sigma_0$, in analogy with the double TlO layer compounds, a $[Tl^{+3}]/[Tl^{+1}]$ ratio of 2 (i.e., $f_{+1} = 1/3$; $f_{+3} = 2/3$) is calculated as the ideally expected value at optimization. Comparing $f_{+1}$ values of three non-optimal $(Tl_{1-x}Cu_x)Ba_2(Ca/Tl)_2Cu_3O_{9\pm\delta}$ compounds with that of the optimal Tl-1223 material confirms that the variation of $f_{+1}$ with $T_C$ trends to 1/3 at optimal composition as predicted, and validating the proposition that $\sigma = \sigma_0$. The fact that $\sigma = \sigma_0$ for ten of the 16 Y-, Tl-, and Hg-based cuprates listed in Table 1, suggests that $\sigma_0$ possesses fundamental significance, as conjectured for a quantum critical point in optimal high-$T_C$ superconductors.

As is shown in Fig. 2, the differences between measured and calculated values of $T_{C0}$ for all 48 compounds, covering seven superconductor families, approaches the normal distribution of error. This analysis supports an absence of bias with regard to the selection of optimal compound parameters. The standard deviation between measured and calculated $T_{C0}$ is 1.31 K with an uncertainty of ± 0.02 K as determined from statistical analysis and is of magnitude similar to earlier findings.

Considered from the perspective of a new Coulombic model for the high-$T_C$ pairing mechanism, the optimal transition temperatures of the seven Tl-based compounds are therefore understood in terms of their respective crystal structure and fractional charge σ, as expressed by Eq. (1), the latter determined from the cation oxidation state and content in both the inner and outer type I layers and the electronegativity of the inner-layer cations. This work completes the first-order determination of the fractional charge values for the $Tl_mBa_2Ca_{\eta-1}Cu_\eta O_{2\eta+2+m}$ (for η = 1, 2, 3; m = 1, 2) series of compounds based on the model of Section 2, and provides important insight into the design of new high-$T_C$ materials.

## Acknowledgements

We are grateful for support from the Physikon Research Corporation (Project No. PL-206), the New Jersey Institute of Technology, and the University of Notre Dame. We also thank D. B. Mitzi for helpful discussions. This work has been published [72].

## References

[1] A. E. Schlögl, J. J. Neumeier, J. Diederichs, C. Allgeier, J. S. Schilling, and W. Yelon, Physica C 216 (1993) 417.

[2] T. Machida, Y. Kamijo, K. Harada, T. Noguchi, R. Saito, T. Kato, and H. J. Sakata, J. Phys. Soc. Jpn. 75 (2006) 083708.

[3] J. Meng, G. Liu, W. Zhang, L. Zhao, H. Liu, W. Lu, X. Dong, and X. J. Zhou, Supercond. Sci. Technol. 22 (2009) 045010.

[4] C. C. Torardi, M. A. Subramanian, J. C. Calabrese, J. Gopalakrishnan, E. M. McCarron, K. J. Morrissey, T. R. Askew, R. B. Flippen, U. Chowdhry, and A. W. Sleight, Phys. Rev. B 38 (1988) 225.

[5] H. Eisaki, N. Kaneko, D. L. Feng, A. Damascelli, P. K. Mang, K. M. Shen, Z.-X. Shen, and M. Greven, Phys. Rev. B 69 (2004) 064512.

[6] M. A. Subramanian, J. C. Calabrese, C. C. Torardi, J. Gopalakrishnan, T. R. Askew, R. B. Flippen, K. J. Morrissey, U. Chowdhry, and A. W. Sleight, Nature 332 (1988) 420.

[7] D. R. Harshman, A. T. Fiory, and J. D. Dow, J. Phys.: Condens. Matter **23** (2011a) 295701; D. R. Harshman, A. T. Fiory, and J. D. Dow, J. Phys: Condens. Matter 23 (2011) 349501, Corrigendum.

[8] D. R. Harshman and A. T. Fiory, J. Phys.: Condens. Matter 24 (2012) 135701.

[9] D. R. Harshman and A. T. Fiory, Phys. Rev. B 86 (2012) 144533.

[10] T. Manako and Y. Kubo, Phys. Rev. B 50 (1994) 6402.

[11] T. Mochiku, T. Nagashima, M. Watahiki, Y. Fukai, and H. Asano, Jpn. J. Appl. Phys. 28 (1989) L1926.

[12] E. Abrahams, Int. J. Mod. Phys. 24 (2010) 4150.

[13] P. M. C. Rourke, A. F. Bangura, T. M. Benseman, M. Matusiak, J. R. Cooper, A. Carrington, and N. E. Hussey, New J. Phys. 12 (2010) 105009.

[14] J. Zaanen, S. Chakravarty, T. Senthil, P. Anderson, P. Lee, J. Schmalian, M. Imada, D. Pines, M. Randeria, C. Varma, M. Vojta, and M. Rice, Nat. Phys. 2 (2006) 138.

[15] C. M. Varma, Rep. Prog. Phys. 75 (2012) 052501.

[16] V. Aji, A. Shekhter, and C. M. Varma, Phys. Rev. B 81 (2010) 064515.

[17] W. A. Little, Novel Superconductivity, in: S. A. Wolf, V. Z. Kresin (Eds.), 1988, Plenum, New York, p. 341.

[18] A. Bill, H. Morawitz, and V. Z. Kresin, Phys. Rev. B 68 (2003) 144519.

[19] R. Wang, H. Zhang, L. Zhang, and H. Shi, Tsinghua Sci. Tech. 1 (1996) 253.

[20] Q. Luo and R. Wang, J. Phys. Chem. Solids 48 (1987) 425.

[21] R. Jayaprakash and J. Shanker, J. Phys. Chem. Solids 54 (1993) 365.

[22] S. Balasubramanian and K. J. Rao, Solid State Commun. 71 (1989) 979.

[23] K. Saito and M. Kaise, Phys. Rev. B 57 (1998) 11786.

[24] M. Z. Cieplak, S. Guha, S. Vadlamannati, T. Giebultowicz, and P. Lindenfeld, Phys. Rev. B 50 (1994) 12876.

[25] D. R. Harshman and A. T. Fiory, Emerg. Mater. Res. 1 (2012) 4.

[26] P. J. Baker, T. Lancaster, S. J. Blundell, F. L. Pratt, M. L. Brooks, and S.-J. Kwon, Phys. Rev. Lett. 102 (2009) 087002.

[27] D. R. Harshman and A. T. Fiory, Phys. Rev. B 90 (2014) 186501.

[28] D. R. Harshman, A. T. Fiory, J. Supercond. Nov. Mater. (2015), Submitted for publication.

[29] C. Boulesteix, Y. Marietti, T. Badèche, H. Tatarenko-Zapolshy, V. Grachev, O. Monnereau, H. Faqir, and G. Vacquier, J. Phys. Chem. Solids 61 (1999) 585.

[30] W. Fan and Z. Zeng, Supercond. Sci. Technol. 24 (2011) 105007.

[31] G. D. Gu, R. Puzniak, K. Nakao, G. J. Russell, and N Koshizuka, Supercond. Sci. Technol. 11 (1998) 1115.

[32] D. E. Farrell, S. Bonham, J. Foster, Y. C. Chang, P. Z. Jiang, K. G. Vandervoort, D. J. Lam, and V. G. Kogan, Phys. Rev. Lett. 63 (1989) 782.

[33] J. T. Moonen, D. Reefman, J. C. Jol, H. B. Brom, T. Zetterer, D. Hahn, H. H. Otto, and K. F. Renk, Physica C 185 (1991) 1891.

[34] G. G. Li, J. Mustre de Leon, S. D. Conradson, M. V. Lovato, and M. A. Subramanian, Phys. Rev. B 50 (1994) 3356.

[35] G. G. Li, F. Bridges, J. B. Boyce, T. Claeson, C. Strom, S.-G. Eriksson, and S. D. Conradson, Phys. Rev. B 51 (1995) 8564.

[36] N. Winzek, F. Hentsch, M. Mehring, Hj. Mattausch, R. Kremer, and A. Simon, Physica C 168 (1990) 327.

[37] P. G. Radaelli, D. G. Hicks, A. W. Mitchell, B. A. Hunter, J. L. Wagner, B. Dabrowski, K. G. Vandervoort, H. K. Viswanathan, and J. D. Jorgensen, Phys Rev. B 49 (1994) 4163.

[38] M. A. Subramanian, G. H. Kwei, J. B. Parise, J. A. Goldstone, and R. B. Von Dreele, Physica C 166 (1990) 19.

[39] H. Ihara and Y. Sekita, U.S. Patent No. 6,300,284 (Oct. 9, 2001).

[40] A. Sundaresan, C. S. Gopinath, S. Subramanian, L. C. Gupta, M. Sharon, R. Pinto, and R. Vijayaraghavan, Phys. Rev. B 50 (1994) 10238.

[41] A. Sundaresan, C. S. Gopinath, A. S. Tamhane, A. K. Rajarajan, M. Sharon, S. Subramanian, P. Pinto, L. C. Gupta, and R. Vijayaraghavan, Phys. Rev. B 46 (1992) 6622.

[42] H. H. Otto, R. Baltrusch, and H.-J. Brandt, Physica C 215 (1993) 205.

[43] F. Hentsch, N. Winzek, M. Mehring, Hj. Mattausch, and A. Simon, Physica C 158 (1989) 137.

[44] B. Morosin, E. L. Venturini, and D. S. Ginley, Physica C 175 (1991) 241.

[45] B. Morosin, E. L. Venturini, and D. S. Ginley, Physica C 183 (1991) 90.

[46] M. A. Subramanian, J. B. Parise, J. C. Calabrese, C. C. Torardi, J. Gopalakrishnan, and A. W. Sleight, J. Sol. State Chem. 77 (1988) 192.

[47] K. Tanaka, A. Iyo, N. Terada, K. Tokiwa, S. Miyashita, Y. Tanaka, T. Tsukamoto, S. K. Agarwal, T. Watanabe, and H. Ihara, Phys. Rev. B 63 (2001) 064508.

[48] N. Terada, K. Tanaka, Y. Tanaka, A. Iyo, K. Tokiwa, T. Watanabe, and H. Ihara, Physica B 284-288 (2000) 1083.

[49] A. Iyo, Y. Tanaka, Y. Ishiura, M. Tokumoto, K. Tokiwa, T. Watanabe, and H. Ihara, Supercond. Sci. Technol. 14 (2001) 504.

[50] K. Maiti, J. Fink, S. de Jong, M. Gorgoi, C. Lin, M. Raichle, V. Hinkov, M. Lambacher, A. Erb, and M. S. Golden, Phys. Rev. B 80 (2009) 165132.

[51] D. A. Nepela and J. M. McKay, Physica C 158 (1989) 65.

[52] C. M. Varma, Z. Nussinov, and W. van Saarloos, Phys. Rep. 361 (2002) 267.

[53] S. Sachdev, Rev. Mod. Phys. 75 (2003) 913.

[54] D. N. Basov and T. Timusk, Rev. Mod. Phys. 77 (2005) 721.

[55] C. M. Varma, Nature 468 (2010) 184.

[56] A. P. Mackensie, S. R. Julian, D. C. Sinclair, and C. T. Lin, Phys. Rev. B 53 (1996) 5848.

[57] M. Merz, N. Nücker, P. Schweiss, S. Schuppler, C. T. Chen, V. Chakarian, J. Freeland, Y. U. Idzerda, M. Kläser, G. Müller-Vogt, and Th. Wolf, Phys. Rev. Lett. 80 (1998) 5192.

[58] D. R. Harshman, J. D. Dow, and A. T. Fiory, Philos. Mag. 91 (2011) 818.

[59] Y. Kubo and T. Manako, Physica C 197 (1992) 378.

[60] P. S. Wang, J. C. Williams, K. D. D. Rathnayaka, B. D. Hennings, D. G. Naugle, and A. B. Kaiser, Phys. Rev. B 47 (1993) 1119.

[61] N. Nücker, E. Pellegrin, P. Schweiss, J. Fink, S. L. Molodtsov, C. T. Simmons, G. Kaindl, W. Frentrup, A. Erb, and G. Müller-Vogt, Phys. Rev. B 51 (1995) 8529.

[62] T. Manako, Y. Shimakawa, Y. Kubo, T. Satoh, and H. Igarashi, Physica C 158 (1989) 143.

[63] H. C. Ku, M. F. Tai, J. B. Shi, M. J. Shieh, S. W. Hsu, G. H. Hwang, D. C. Ling, T. J. Watson-Tang, and T. Y. Lin, Jpn. J. Appl. Phys. 28 (1989) L923.

[64] I. K. Gopalakrishnan, J. V. Yakhmi, and R. M. Iyer, Physica C 175 (1991) 183.

[65] P. A. Miles, S. J. Kennedy, G. J. McIntyre, G. D. Gu, G. J. Russell, and N. Koshizuka, Physica C 294 (1998) 275.

[66] F. Munakata, T. Kawano, and H. Yamauchi, J. Sol. State Chem. 101 (1992) 41.

[67] T. H. Geballe, J. Supercond. Nov. Magn. 19 (2006) 261; T. H. Geballe, Annu. Rev. Condens. Matter Phys. 4 (2013) 1.

[68] S. A. Wolf and V. Z. Kresin, J. Supercond. Nov. Magn. 25 (2012) 175.

[69] M. R. Beasley, J. E. Mooij, and T. P. Orlando, Phys. Rev. Lett. 42 (1979) 1165.

[70] D. R. Harshman and A. T. Fiory, U.S. Patent No. 8,703,651 (22 April 2014).

[71] O. Gunnarsson, Rev. Mod. Phys. 69 (1997) 575.

[72] D. R. Harshman and A. T. Fiory, J. Phys. Chem. Solids 85 (2015)106-116.